# Surface-diffusion-limited growth of atomically thin $WS_2$ crystals from core-shell nuclei

Sunghwan Jo[1], Jin-Woo Jung[1], Jaeyoung Baik[1], Jang-Won Kang[1], Il-Kyu Park[2], Tae-Sung Bae[3], Hee-Suk Chung[3]*, and Chang-Hee Cho[1]*

[1]Department of Emerging Materials Science, Daegu Gyeongbuk Institute of Science and Technology (DGIST), Daegu 42988, South Korea

[2]Department of Materials Science and Engineering, Seoul National University of Science and Technology, Seoul 139-743, South Korea

[3]Analytical Research Division, Korea Basic Science Institute, Jeonju 54907, South Korea

*E-mail: chcho@dgist.ac.kr (C.H.C.) and hschung13@kbsi.re.kr (H.S.C)

**Atomically thin transition metal dichalcogenides (TMDs) have recently attracted great attention since the unique and fascinating physical properties have been found in various TMDs, implying potential applications in next-generation devices. The progress towards developing new functional and high-performance devices based on TMDs, however, is limited by the difficulty of producing large-area monolayer TMDs due to a lack of knowledge of the growth processes of monolayer TMDs. In this work, we have investigated the growth processes of monolayer $WS_2$ crystals using a thermal chemical vapor deposition method, in which the growth conditions were adjusted in a systematic manner. It was found that, after forming $WO_3$-$WS_2$ core-shell nanoparticles as nucleation sites on a substrate, the growth of three-dimensional $WS_2$ islands proceeds by ripening and crystallization processes. Lateral growth of monolayer $WS_2$ crystals subsequently occurs by surface diffusion process of adatoms. Our results provide understanding of the growth processes of monolayer $WS_2$ by using chemical vapor deposition methods.**

Two-dimensional (2D) semiconducting layered materials, which are known as transition metal dichalcogenides (TMDs), have received great attention due not only to the unique physical properties[1-4] but also to their potential application in future electronic and optical devices using 2D layered heterostructures.[5-8] Notably, the fascinating properties of TMDs, such as indirect-to-direct bandgap transition in monolayer limit,[1,9] strong excitonic effects,[3,10,11] and valley degrees of freedom,[4,12-15] have stimulated intense research on a variety of 2D TMD materials. In an attempt to examine useful applications, there has been much effort to demonstrate electronic and optoelectronic devices based on 2D TMDs, including field-effect transistors,[16,17] light-emitting diodes[18,19] and lasers.[20,21] However, progress towards developing high-performance devices using 2D TMDs has been greatly challenged by the difficulty of producing large-area 2D TMD thin films.

Although monolayer or few layer TMDs can be obtained by mechanical or chemical exfoliation from bulk crystals,[22,23] these methods cannot be readily employed in conventional thin-film-based device fabrication processes. Since vacuum deposition techniques can provide suitable routes for the growth of TMD thin films, various techniques such as thermal chemical vapor deposition (CVD), [24-27] metal-organic CVD,[28-29] and molecular beam epitaxy[30] have been utilized to grow 2D TMDs on substrates. Among the growth techniques, the thermal CVD method using powder sources has been most commonly used to produce 2D TMDs. Through transmission electron microscopy analysis, a recent study has reported that monolayer growth of $MoS_2$-$MoSe_2$ alloys proceeds from the heterogeneous nuclei of core-shell structured nanoparticles.[31] However, the growth mechanisms of 2D TMDs have still been largely unexplored in terms of observing the 2D growth evolution of monolayer TMDs, thus hampering the large-scale growth of high-quality 2D TMDs and the development of high-performance devices based on 2D TMDs.

In this work, we report on the growth evolution of monolayer $WS_2$ crystals from $WO_3$-$WS_2$ core-shell nuclei based on controlled CVD methods. Transmission electron microscopy (TEM) analysis revealed that one or a few nanoparticles of a $WO_3$-$WS_2$ core-shell structure exist in the CVD-grown $WS_2$ crystals, acting as nucleation sites in the initial stage of crystal growth. A systematic study performed by varying the growth conditions further revealed that after forming $WO_3$-$WS_2$ core-shell nanoparticles on the substrate, $WS_2$ 3D islands grow from the core-shell nanoparticles by crystallization and ripening processes, and subsequently, lateral growth of 2D monolayer $WS_2$ crystal occurs by surface diffusion of W-S molecular clusters, generating a maximal size of the 2D monolayer domain due to the limited length scale of surface diffusion and the limited number of W-S molecular clusters. Our work offers insight into the growth mechanisms of monolayer TMDs using CVD methods.

Monolayer $WS_2$ crystals were grown using a CVD system, which enables control of the growth conditions in a precise manner (Supplementary Information Figure S1a). Ten milligrams of $WO_3$ powder in a quartz boat was placed in the center of the CVD chamber, and 400 mg of S powder in a quartz boat was placed upstream, 5 cm away from the $WO_3$ boat. A 270-nm-thick $SiO_2$-coated Si wafer was used as a growth substrate, which was spin-coated with a seed promoter layer of perylene-3,4,9,10-tetracarboxylic acid tetrapotassium salt (PTAS).[32] The growth substrate was placed face-down above the quartz boat containing $WO_3$ powder. The furnace was ramped from room temperature to the growth temperature, held for a specified holding time, and finally cooled down to room temperature (Supplementary Information Figure S1b). During growth, the chamber pressure was kept at 600 Torr in an Ar flow of 50 sccm, while the growth temperature and time were varied to study the growth mechanism of monolayer $WS_2$. Figure 1a presents an optical microscope image of $WS_2$ crystal domains grown at a growth temperature of 850 °C for a hold time of 20 min. For most triangular $WS_2$ crystals, 3-dimensional (3D) islands are consistently observed at the center of each $WS_2$ crystal. Figure 1b shows the micro-photoluminescence spectrum measured with a 457.9 nm Ar ion laser excitation source at room temperature. Strong photoluminescence was observed on the 2-dimensional (2D) $WS_2$ crystals, and the photoluminescence peaked at a wavelength of ~630 nm, corresponding to the neutral and charged states of the A-exciton of monolayer $WS_2$. This finding indicates that 2D $WS_2$ sheets are present in the form of monolayers because a direct to indirect bandgap transition appears when increasing the number of layers from a monolayer, leading to drastic quenching of the photoluminescence intensity.[1,9] Figure 1c presents the annular dark field (ADF)-scanning transmission electron microscopy (STEM) image of a $WS_2$ crystal, which was transferred onto a carbon-coated copper grid for TEM sample preparation. It was confirmed that one or a few nanoparticles exist mostly in the 3D island domains. As shown in Figure 1d, the nanoparticles possess a core-shell structure, suggesting the role of core-shell nanoparticles as nucleation sites in the growth of $WS_2$ crystals. The details of the TEM analysis are discussed later.

In order to investigate the structural properties of as-grown $WS_2$, micro-Raman spectroscopy and atomic force microscopy (AFM) were carried out for both the 3D island and the 2D sheet regions, as indicated in the optical microscope image of Figure 2a. Figure 2b presents micro-Raman spectra, showing the $E^1_{2g}$ and $A_{1g}$ modes of $WS_2$ measured at the 3D island and 2D sheet regions. The measured spectra indicate that the 3D island and the 2D sheet regions possess the same crystalline symmetry since $WS_2$ crystals belong to the space group $D_{6h}$ ($P6_3/mmc$), which has two Raman-active modes of $E^1_{2g}$ and $A_{1g}$.[33,34] The only difference between the 3D island and the 2D sheet is the peak separation between the $E^1_{2g}$ and $A_{1g}$ modes, which was measured to be 63.8 ($\Delta\omega_{3D}$) and 60.9 ($\Delta\omega_{2D}$) cm$^{-1}$, respectively. It is well known that peak separation increases with an increasing number of $WS_2$ layers because the $E^1_{2g}$ mode softens, while the $A_{1g}$ mode hardens due to interlayer interactions when increasing the number of layers.[35] The measured peak separation value for the 2D sheet is in good agreement with previously reported values for monolayer $WS_2$, implying that the 2D sheet region has the thickness of monolayer $WS_2$.[35,36] Figure 2c presents the AFM topography, which clearly shows the 3D island and 2D monolayer regions of the as-grown $WS_2$ crystal. In addition, the thickness was measured using the AFM step height profiles, as shown in Figure 2d. The thickness of the 2D monolayer $WS_2$ is estimated to be 0.9 nm, which is consistent with the reported values,[35,36] while that of the 3D island is as high as 42 nm, implying the existence of embedded core-shell nanoparticles, as observed in Figure 1d.

The atomic scale structure and chemical composition of the 3D island and 2D $WS_2$ crystal were studied using TEM techniques, which can elucidate the role of the 3D island in the growth of 2D $WS_2$ crystals. Figure 3a displays the bright-field (BF) TEM image showing a 3D island and a 2D $WS_2$ crystal at low magnification. To investigate the crystallinity of the $WS_2$ domain, selected area electron diffraction (SAED) patterns were obtained at the center (red square) and the outer region (blue square) of the triangular $WS_2$ crystal. The insets of Figure 3a show identical single crystallinity at the 3D island and 2D $WS_2$ sheet regions, which is consistent with the Raman analysis result in Figure 2b. Interestingly, an ellipsoidal nanoparticle was found in the 3D island $WS_2$ area (red square). For all the $WS_2$ crystals seen in the TEM image, one or a few nanoparticles are consistently observed mostly in the 3D core region. This result strongly suggests that the nanoparticles act as nucleation sites in the initial stage of 3D island $WS_2$ growth. Figure 3b presents a magnified ADF-STEM image of the nanoparticle observed in Figure 3a. The nanoparticle shows a core-shell structure with a core size of approximately 100 nm and a shell thickness of approximately 20 nm. To investigate the structural and chemical information, high-resolution (HR)-ADF-STEM was performed for the core-shell interface (green square) and the 2D sheet region (yellow square) in Figure 3b. The HR-ADF-STEM image at the core-shell interface is shown in Figure 3c. For the core region, the atomic-resolved $WO_3$ single crystalline structure was clearly seen in the [011] zone axis, as shown in the inset of Figure 3c. In the shell region, a typical layered structure of single-crystalline $WS_2$ was found with a $WS_2$ (001) d-spacing of 6.5 Å.[31,37] In addition, energy dispersive X-ray spectrometry (EDS) mapping of the core-shell nanoparticle provides clear spatial and chemical information of the tungsten, sulfur and oxygen atoms (Supplementary Information Figure S2). Figure 3d offers an HR-ADF-STEM image of the 2D sheet region (yellow square in Figure 3b), which shows single-crystalline $WS_2$ with a 2H structure. A fast Fourier transform (FFT) pattern and enlarged images (the insets of Figure 3d) also provide clear evidence of the 2H single-crystalline $WS_2$ structure.[38] These TEM results reveal that $WO_3$-$WS_2$ core-shell nanoparticles are consistently present and may act as nucleation sites in the initial stage of crystal growth.

To further investigate the layered structure and chemical properties of the $WS_2$ domain, cross-sectional TEM and EDS mapping were performed with a sample prepared by a focused ion beam (FIB) lift-out technique. Figure 3e displays a low magnification BF TEM image showing the cross-sectional view of the $WS_2$ crystal. A stepped $WS_2$ image confirms the cross-sectional structure from the 3D island to 2D $WS_2$ crystal regions. As shown in Figure 3f, the cross-sectional HR-TEM image of $WS_2$ reveals a typical layered $WS_2$ structure, as confirmed in the $WS_2$ shell region from the plane view TEM analysis. EDS mapping on the ADF-STEM image confirms the well-defined spatial elemental distribution for tungsten, sulfur and oxygen atoms (Figure 3g).

To understand the details of the growth process, the morphological evolution was monitored while varying the growth conditions of temperature and hold time. Figure 4a shows the optical microscope images of $WS_2$ grown at the temperatures of 500, 750, and 850 °C for a constant hold time of 0 min. At the low temperature of 500 °C, a high density of nanoparticles is observed, and they are most likely the nucleation centers containing $WO_3$-$WS_2$ core-shell nanoparticles, as confirmed in the TEM analysis. As the growth temperature increases to 750 °C, small $WS_2$ domains (~a few μm in size) start to appear mostly in the form of 3D islands. As the temperature is further increased to 850 °C, the size of the $WS_2$ domains is increased to ~10 μm with an enlarged 2D monolayer region while remaining the similar size of 3D islands. These results indicate that ramping the temperature during the growth process causes the formation of nucleation centers at low temperature, which are facilitated by the $WO_3$-$WS_2$ nanoparticles, and, through the formation of 3D island domains at the intermediate temperature, 2D monolayer growth proceeds at the growth temperature of 850 °C. Importantly, it should be noted that the S powder in the growth chamber is fully evaporated and exhausted from the source boat at the temperature of 500 °C during the ramping process (Supplementary Information Figure S3). This suggests a scenario for the growth procedure as follows. In the early stage, a high density of $WO_3$-$WS_2$ core-shell nanoparticles form as nucleation centers on the PTAS-coated substrate during the ramping time. As the temperature increases, $WS_2$ 3D islands are grown without supplying the vapor phase S source due to the exhaustion of S powder, and the lateral growth of a 2D monolayer begins most likely due to the surface migration and coalescence of W-S molecular clusters through a 2D ripening effect.[39-41]

Figure 4b shows the evolution of $WS_2$ 2D growth as the hold time is increased from 0 to 150 min at a temperature of 850 °C. For a hold time of 20 min, the size of the triangular $WS_2$ 2D domain reaches a value larger than 20 μm, indicating that growth of a 2D monolayer is further facilitated with a hold time of 20 min. For a hold time longer than 20 min, the size of the 2D monolayer region decreases, but that of the 3D island region increases. Figure 4c presents the time-dependent domain size for both the 2D monolayer and 3D island regions at a growth temperature of 850 °C. With an increase in hold time, the size of the 2D monolayer initially increases and becomes greatest with a size of ~20 μm at 20 min and then decreases due to re-evaporation of $WS_2$ from the 2D monolayer region. Meanwhile, the size of the 3D island increases continually, and eventually, only 3D island regions exist without a 2D monolayer region at a hold time of 150 min. These results provide insight into the lateral growth of $WS_2$ 2D domains.

Without supplying the vapor phase S source, lateral 2D growth would proceed by surface diffusion of the W-S molecular clusters (adatoms) because of the concentration gradient between smaller and larger 3D islands, which is known as the Gibbs-Thomson effect.[39-41] The existence of a maximal size of the 2D monolayer domain is attributed to the limited length scale of surface diffusion and the limited number of W-S molecular clusters near the growing $WS_2$ crystals.

Based on the experimental results, we suggest the growth procedure of monolayer $WS_2$ in a thermal CVD system, as schematically shown in Figure 5. The growth process can be classified into three steps, namely, nucleation, 3D island growth, and 2D monolayer growth in chronological order. In the nucleation step, during the ramping time, $WO_3$ and S evaporate into the vapor phase, and W-O molecular clusters condense onto the PTAS-coated substrate, forming a high density of nanoparticles. Subsequently, in the S-rich environment, the nanoparticle surface is then sulfurized to form the $WS_2$ shell layer, generating $WO_3$-$WS_2$ core-shell nanoparticles on the substrate. Recently, the formation of such core-shell nanoparticles has also been observed in other TMDs of the $MoS_2$-$MoSe_2$ alloy.[31] Moreover, the TMD core-shell nanoparticle formation mechanism is well known in the literature.[42-45] Consequently, the $WO_3$-$WS_2$ core-shell nanoparticles act as nucleation centers for the growth of $WS_2$ crystals. When reaching a temperature of ~500 °C, the S powder in the growth chamber is fully evaporated and exhausted from the source boat. With further increases in temperature in the 3D island growth step, the $WS_2$ 3D islands are grown by crystallization and ripening processes,[39-41] in which W-S molecular clusters may diffuse from smaller to larger nanoparticles on the substrate surface. A previous study reported that the formation energy for bulk $MoS_2$, calculated by first principles density functional theory, is lower than that of monolayer $MoS_2$.[46] This finding indicates that a sufficiently high reaction temperature is required for 2D monolayer growth. In the 2D monolayer growth step with reaching the final growth temperature, the lateral growth of the 2D monolayer $WS_2$ crystal occurs due to surface diffusion of W-S molecular clusters (adatoms) through 2D ripening processes. 2D monolayer growth starts from the step edge of 3D single crystalline $WS_2$ islands, which reduces the nucleation barrier and formation energy.[47,48] Due to the limited length scale of surface diffusion and the limited number of W-S molecular clusters near the growing $WS_2$ crystals, the maximal size of the 2D monolayer domain is limited to 10 to 20 μm under a S source deficient condition.

In conclusion, we have investigated the growth processes of monolayer $WS_2$ crystals using thermal CVD methods. For most CVD-grown $WS_2$ crystals, 3D island regions were observed near the center of 2D monolayer $WS_2$ crystals. TEM analysis revealed that one or a few nanoparticles exist in the 3D islands in the form of the $WO_3$-$WS_2$ core-shell structure, and the nanoparticles act as nucleation sites in the initial stage of crystal growth. The detailed study with varying the growth conditions suggests that the growth process of monolayer $WS_2$ crystals can be understood by 3 steps as follows. First, during the ramping time, $WO_3$-$WS_2$ core-shell nanoparticles are formed as nucleation centers on the substrate. Second, with further increases in temperature, the $WS_2$ 3D islands grow from the core-shell nanoparticles by crystallization and ripening processes. Third, upon reaching the final growth temperature, lateral growth of a 2D monolayer $WS_2$ crystal occurs by surface diffusion of W-S molecular clusters through ripening processes on the substrate. Our results may offer insight into the growth mechanisms of monolayer TMDs and, thus, suggest a rational route for growing large-scale layer-controlled TMDs.

## Methods

*Optical measurements*: The optical properties of monolayer $WS_2$ were characterized by micro-photoluminescence and Raman spectroscopic measurements at room temperature using a home-built optical microscope. To excite the $WS_2$ crystals, a continuous-wave argon-ion laser, which was tuned to a wavelength of 457.9 nm, was used with a 60×, 0.70 NA (20×, 0.45 NA) objective (Nikon) in the photoluminescence and Raman spectroscopy measurements. The photoluminescence (Raman) spectra were measured using a spectrometer and a cooled charge-coupled device with a grating of 300 lines/mm (1200 line/mm).

*TEM characterizations*: TEM and ADF-STEM analysis were conducted with a Cs-corrected TEM (JEOL ARM-200F) at an accelerating voltage of 200 kV. ADF-STEM images were obtained with the following parameters: probe current: ~20 pA; convergence semi-angle: 20 mrad; condenser aperture: 30 μm; and inner collection semi-angle of the ADF detector: 90 mrad. ADF-STEM images are filtered using the Wiener method to reduce background noise. A cross-sectional TEM specimen was prepared by the FIB (FEI Quanta 2D) lift-out method after deposition of a protective platinum film. EDS analysis was performed with an EDAX detector (80 T) and software (Oxford Aztec TEM).

## Acknowledgements

This work was supported by the Basic Science Research Program (2016R1A2B4014448) and the Leading Foreign Research Institute Recruitment Program (2018K1A4A3A03075584) through the National Research Foundation of Korea, and by the DGIST R&D Program (18-BT-02, 18-BD-0401) funded by the Ministry of Science and ICT of the Korean Government. H.S.C. acknowledges the National Research Foundation of Korea (NRF) grant funded by the Korean Government (MSIP) (No. 2015R1C1A1A01052727). Sunghwan Jo and Jin-Woo Jung contributed equally to this work.

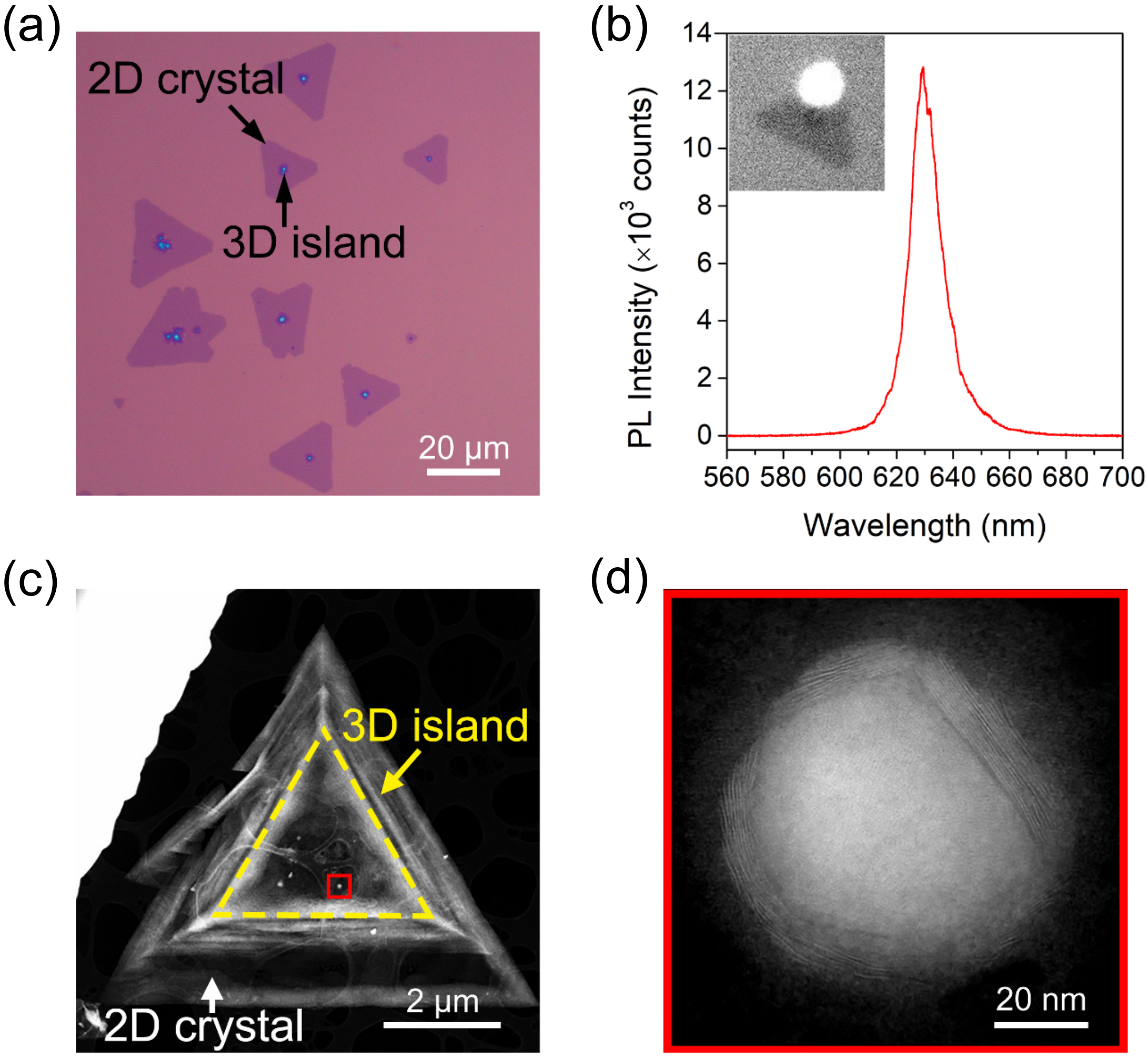

**Figure 1.** (a) Optical microscope image of $WS_2$ crystals grown on a $SiO_2$-coated Si substrate by using thermal chemical vapor deposition. (b) Photoluminescence spectrum taken from the 2D $WS_2$ crystal. Inset: Optical image showing the photoluminescence from the 2D $WS_2$ crystal. (c) Low-magnification ADF-STEM image showing the 3D island and 2D crystal regions of $WS_2$. Note that a large part of the 2D monolayer region is rolled up during the sample transfer process using DI water from the growth substrate to the TEM grid. (d) High-resolution ADF-STEM image of the core-shell nanoparticle marked by the red square in (c).

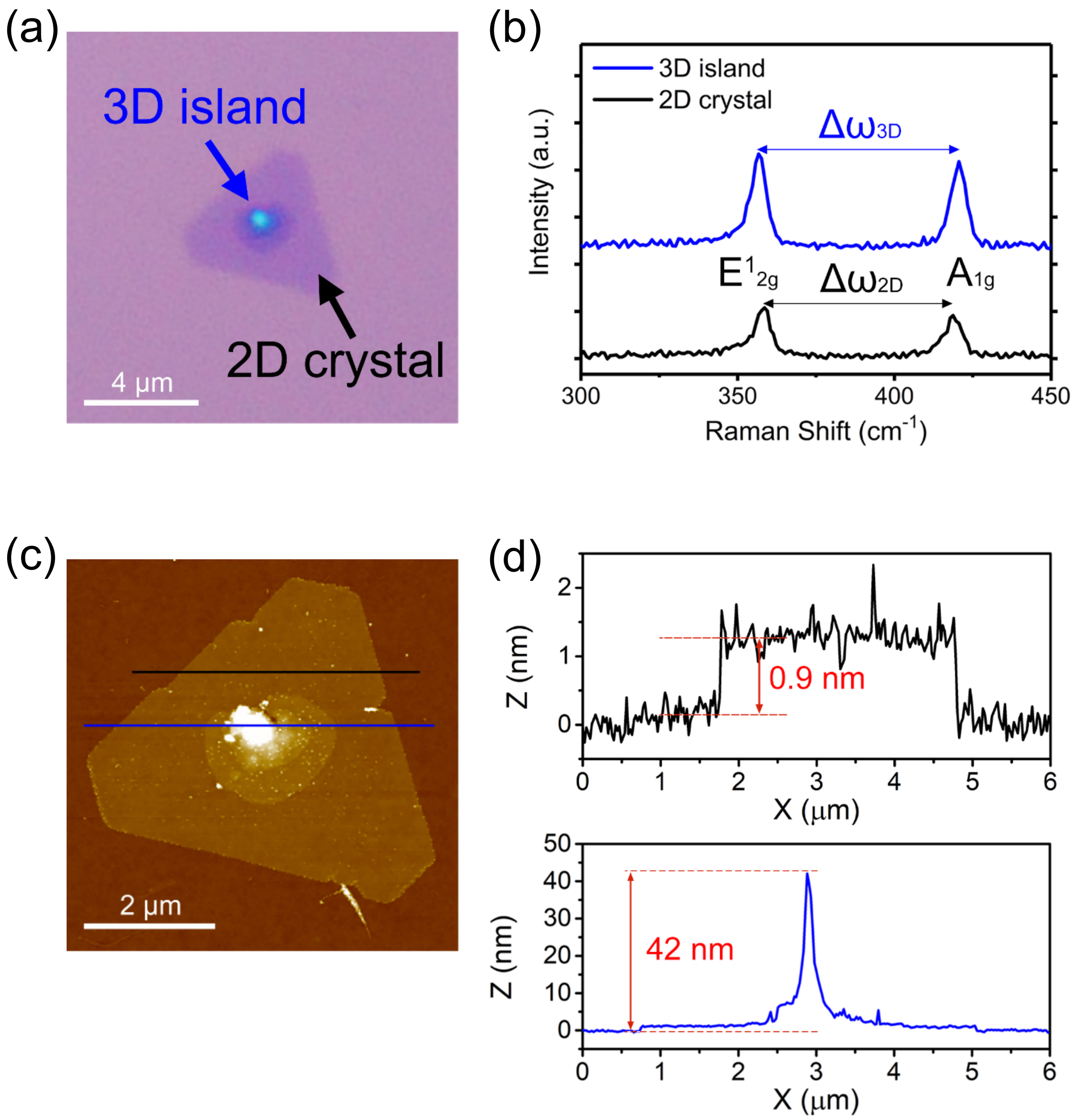

**Figure 2.** (a) Optical microscope image of a $WS_2$ crystal showing the 3D island and 2D crystal regions. (b) Raman spectra obtained from the 3D island (blue) and 2D crystal (black) regions. (c) AFM image of the $WS_2$ crystal. (d) AFM step height profiles along the black and blue lines marked in (c), showing that the thickness of the 2D crystal region is 0.9 nm, corresponding to the monolayer $WS_2$ crystals, while the thickness of the 3D island region is approximately 42 nm, implying the existence of embedded core-shell nanoparticles.

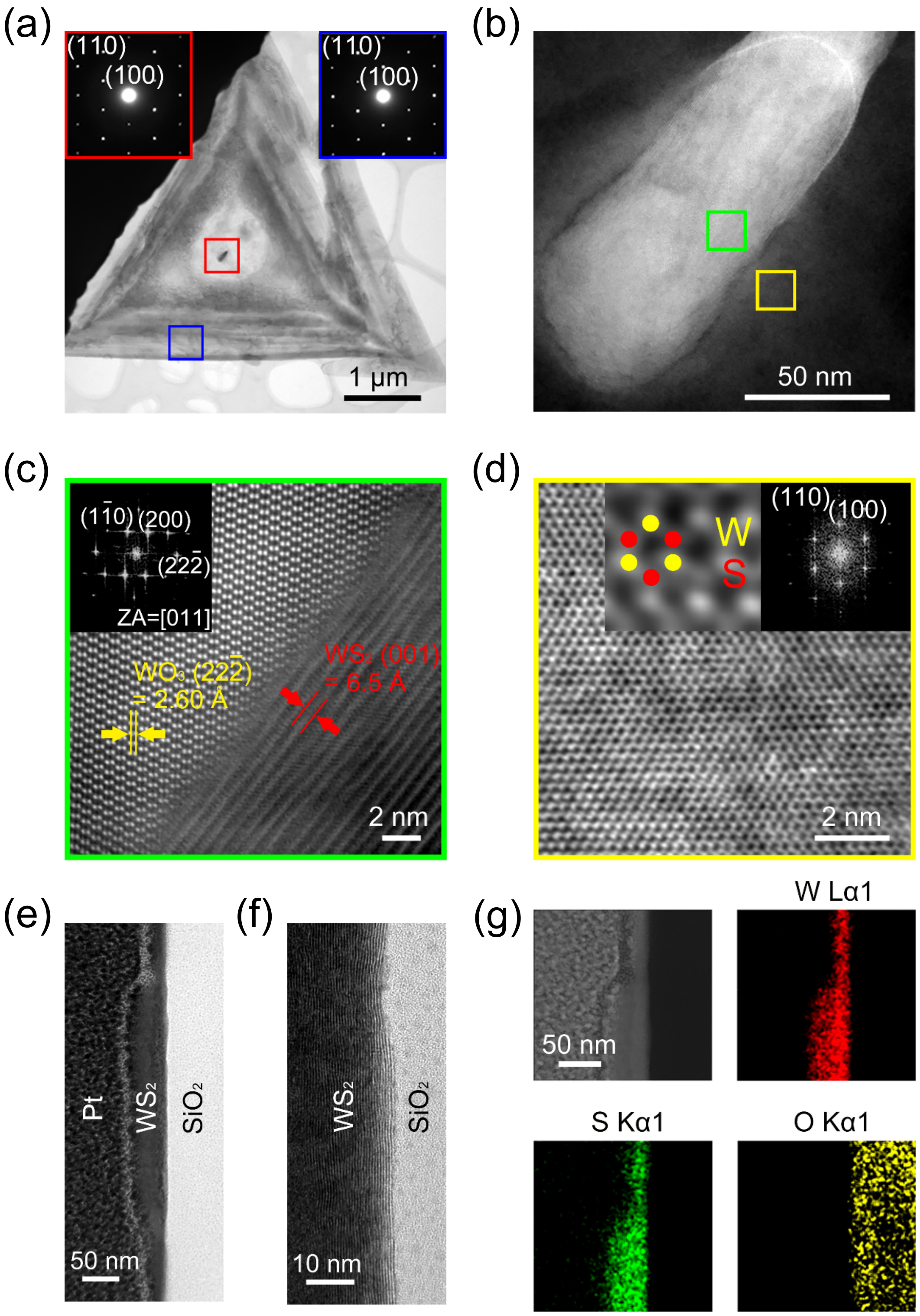

**Figure 3.** (a) Bright-field TEM image of a $WS_2$ crystal. The top-left and top-right insets show the low-magnification SAED patterns obtained from the 3D island and 2D crystal regions of the $WS_2$ crystal marked by red and blue squares on the TEM image. (b) ADF-STEM image of a core-shell nanoparticle in the 3D island (red square region marked in (a)). (c) HR-ADF-STEM image for the core-shell interface region (green square) in (b), and the inset is the FFT pattern obtained from the core region. (d) HR-ADF-STEM image of the 2D crystal region (yellow square) in (b). The left and right insets present a magnified $WS_2$ (2H) HR-ADF-STEM image and FFT pattern of the 2D crystal, respectively. (e) Bright-field TEM image showing the cross-sectional 3D island and 2D crystal regions. (f) HR-TEM image of (e). (g) EDS map images for the cross-sectional 3D island and 2D crystal regions showing the spatial elemental distribution for tungsten (red), sulfur (green) and oxygen (yellow).

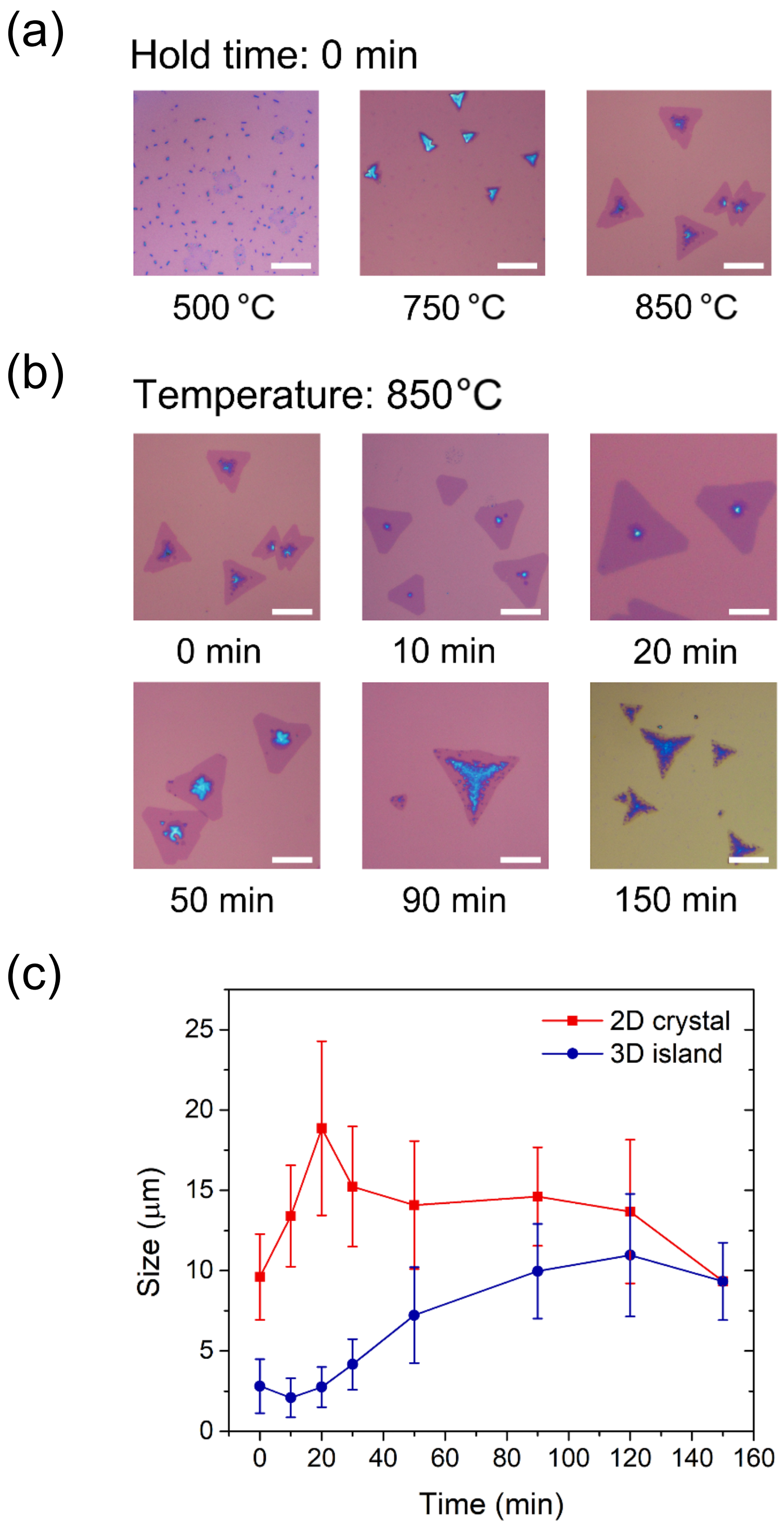

**Figure 4.** (a and b) Optical microscope images showing morphological evolution of CVD-grown $WS_2$ crystal by varying the growth conditions of the temperature (a) and hold time (b). The scale bars are 10

μm. (c) Domain size for the 2D crystal and 3D island regions as a function of the hold time at the temperature of 850 °C.

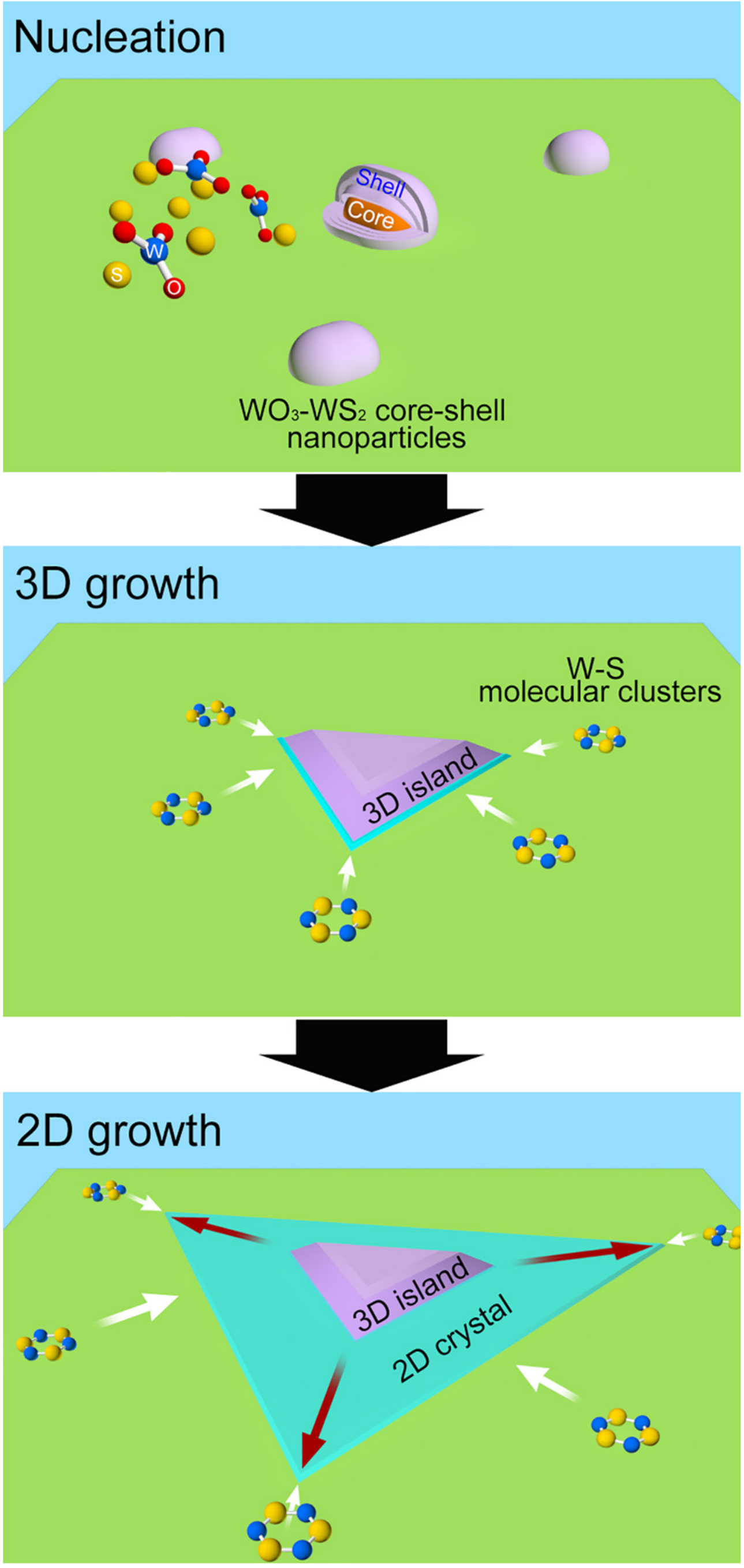

**Figure 5.** Schematic illustration showing the growth processes of $WS_2$ crystals. In the initial stage, $WO_3$-$WS_2$ core-shell nanoparticles are formed on the substrate, and then $WS_2$ 3D islands grow from the core-shell nanoparticles. Finally, the lateral growth of 2D monolayer $WS_2$ occurs by surface diffusion of W-S molecular clusters (adatoms).

Supporting Information

# Surface-diffusion-limited growth of atomically thin $WS_2$ crystals from core-shell nuclei

*Sunghwan Jo, Jin-Woo Jung, Jaeyoung Baik, Jang-Won Kang, Il-Kyu Park, Tae-Sung Bae, Hee-Suk Chung*, and Chang-Hee Cho**

### Experimental setup of the CVD system to grow monolayer $WS_2$

(a)

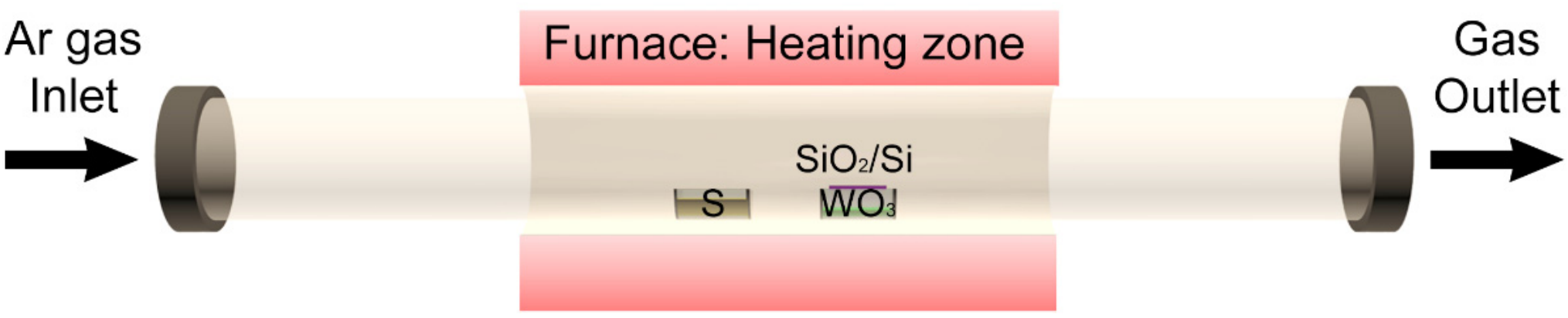

(b)

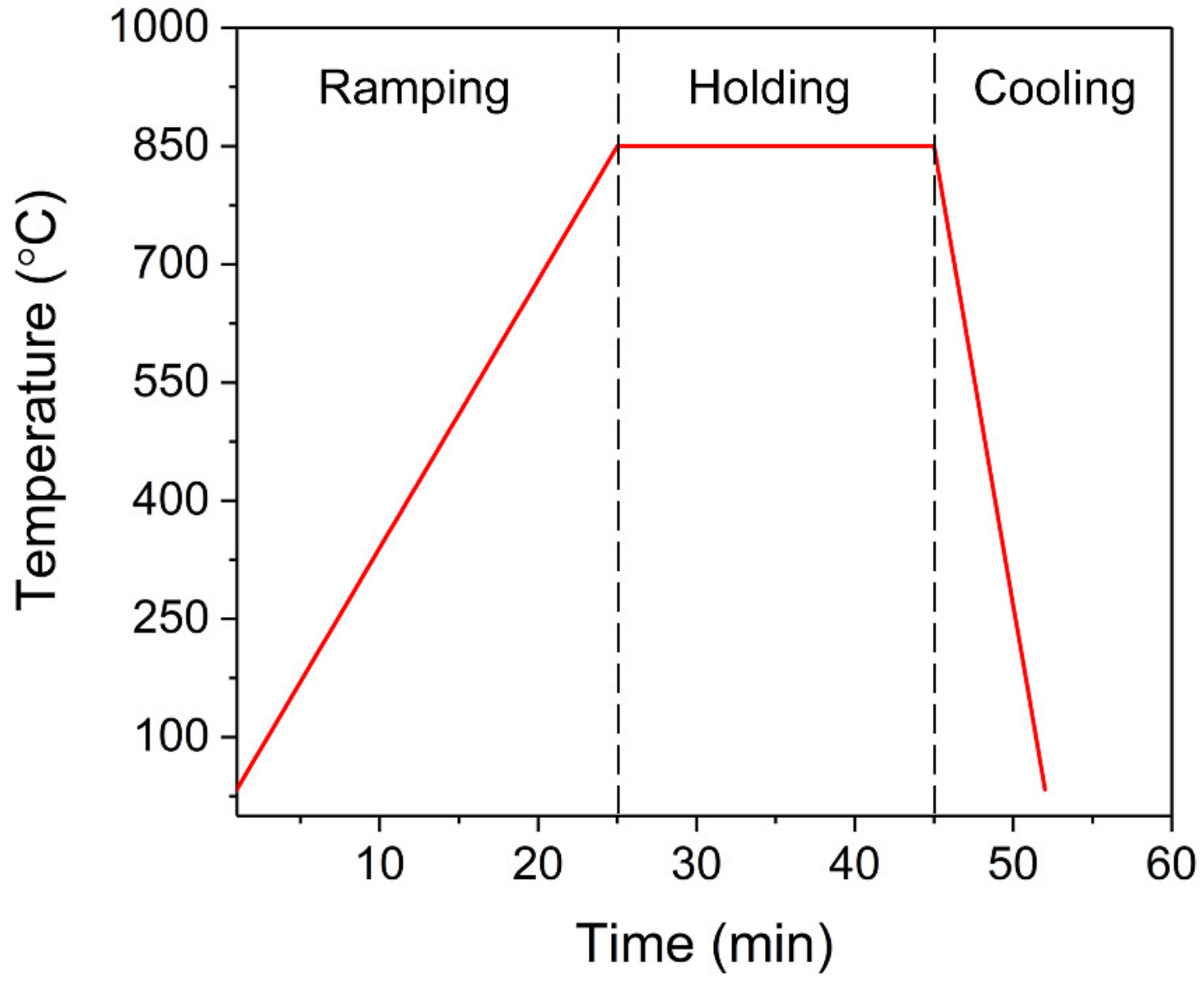

**Figure S1.** (a) Schematic of the thermal CVD system used to synthesize monolayer $WS_2$ monolayer. During the growth process of monolayer $WS_2$, growth conditions, such as the pressure, the temperature and the time, were precisely controlled through the CVD system. (b) The temperature profiles during the growth procedure of monolayer $WS_2$ in the heating zone of the furnace. The furnace was ramped from room temperature to a growth temperature of 850 °C at a rate of 33 °C/min, and the furnace was held at a constant temperature for 20 min and finally cooled down to room temperature.

## EDS map of a core-shell nanoparticle

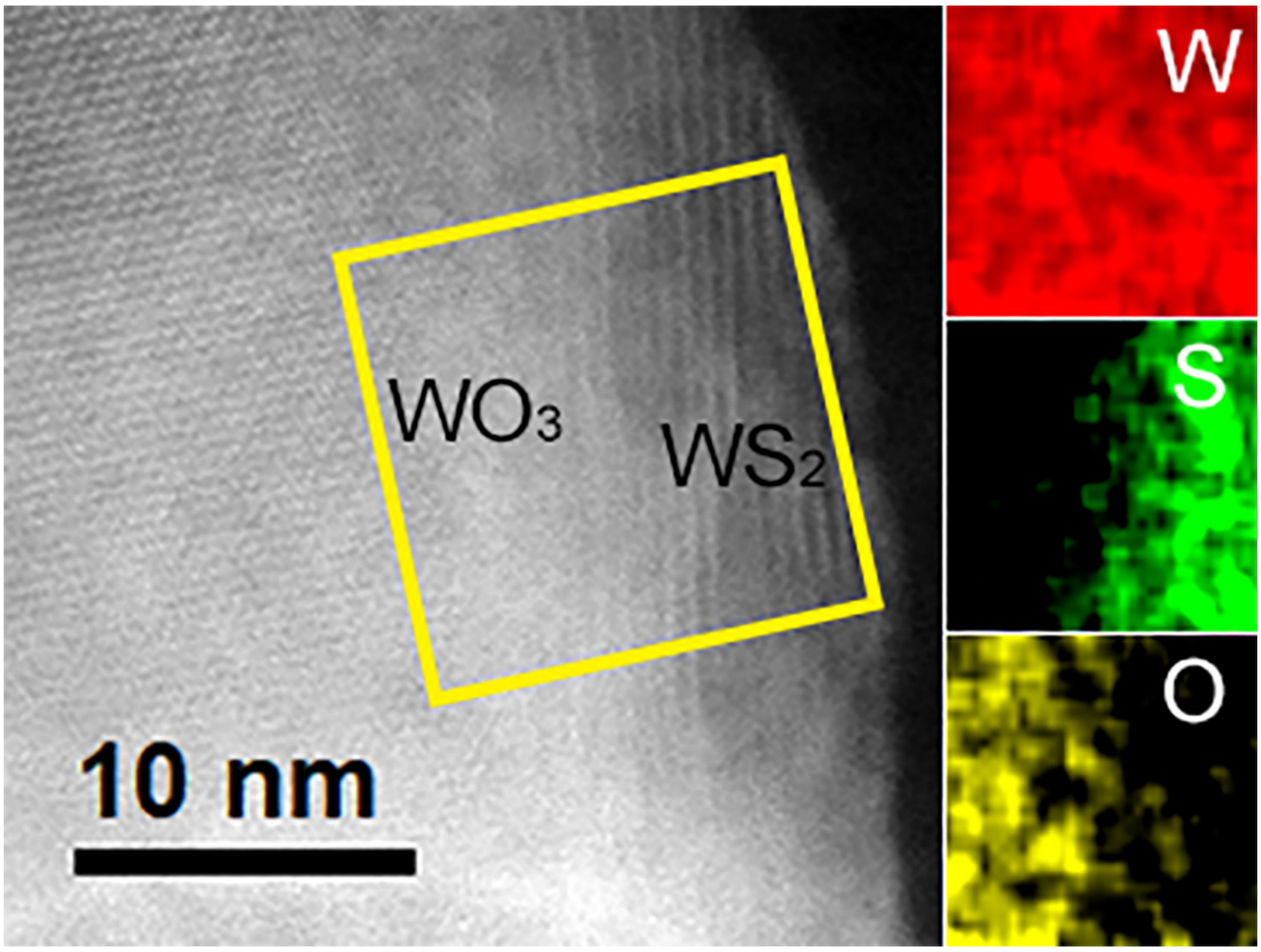

**Figure S2.** ADF-STEM image corresponding to the EDS map of a core-shell nanoparticle in a 3D island $WS_2$ crystal. EDS mapping of the core-shell nanoparticle was performed for the yellow square region in the ADF-STEM image. This EDS map reveals that the core-shell nanoparticle is composed of W, S and O atoms with clear spatial and chemical information.

## Exhaustion of sulfur powder during the ramping period

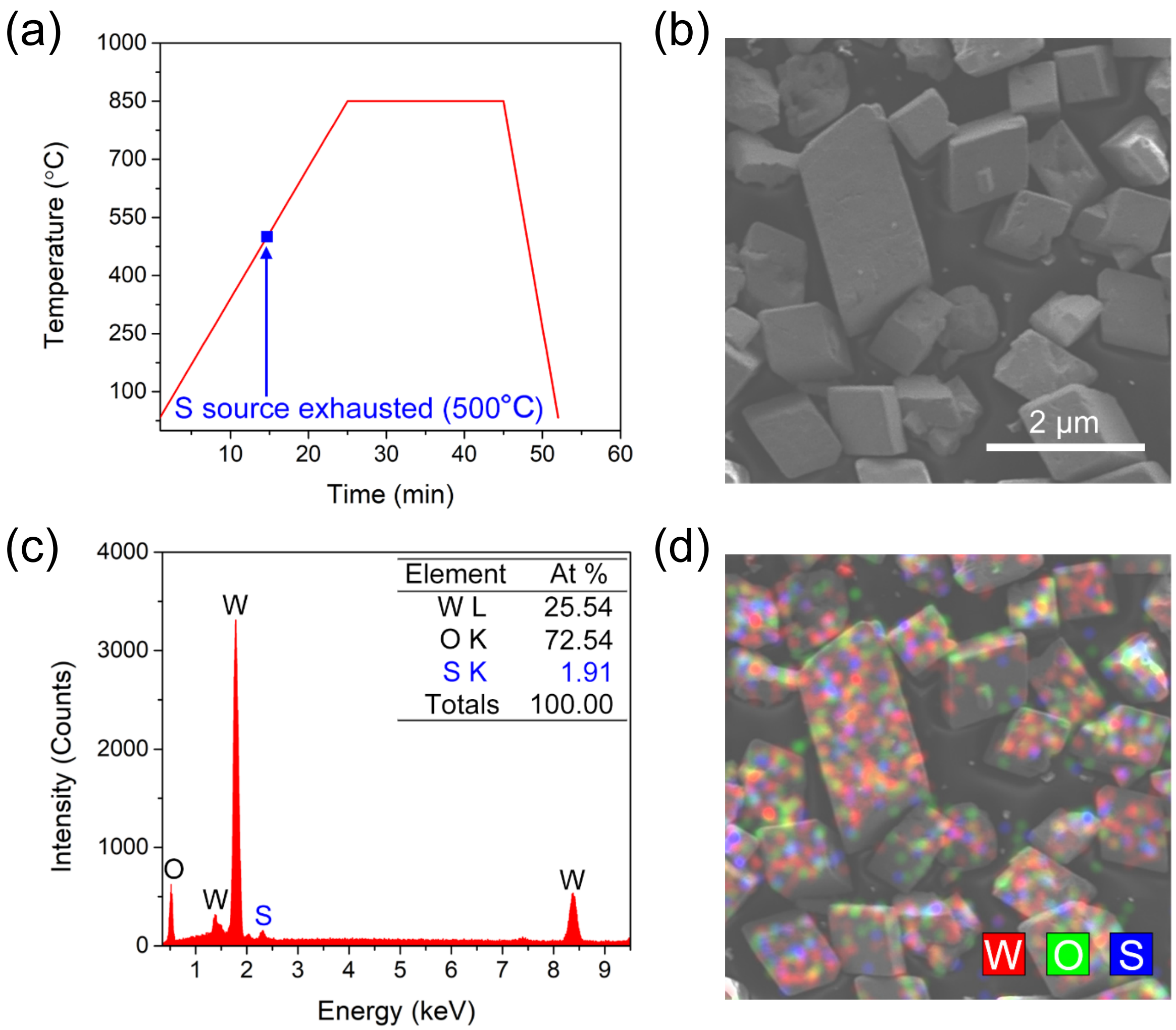

**Figure S3.** (a) Temperature profiles for the growth of monolayer $WS_2$ in the heating zone of the furnace. However, S powder was fully evaporated and exhausted at the temperature of 500 °C before the growth of monolayer $WS_2$. The blue square point indicates the point at which the S powder is fully exhausted from the growth chamber. Note that full evaporation of S powder at the temperature of 500 °C was confirmed by stopping the growth process under the same growth conditions. Additionally, SEM-EDS elemental mapping was performed to check the remaining S element in the $WO_3$ powder, which was heated in the growth chamber, as shown in the following figures. (b) SEM image of the $WO_3$ powder heated to the blue square point in (a). (c, d) EDS spectrum (c) and mapping image (d) of W, O and S elements taken for the full area of the SEM image in (b).